# *Attack resilient architecture to replace embedded Flash with STTRAM in homogeneous IoTs*

Asmit De, Mohammad Nasim Imtiaz Khan and Swaroop Ghosh
Computer Science and Engineering, University of South Florida
{asmitde, khan12, swaroopghosh}@mail.usf.edu

*Abstract*—Spin-Transfer Torque RAM (STTRAM) is an emerging Non-Volatile Memory (NVM) technology that provides better endurance, write energy and performance than traditional NVM technologies such as Flash. In embedded application such as microcontroller SoC of Internet of Things (IoT), embedded Flash (eFlash) is widely employed. However, eFlash is also associated with cost. Therefore, replacing eFlash with STTRAM is desirable in IoTs for power-efficiency. Although promising, STTRAM brings several new security and privacy challenges that pose a significant threat to sensitive data in memory. This is inevitable due to the underlying dependency of this memory technology on environmental parameters such as temperature and magnetic fields that can be exploited by an adversary to tamper with the program and data. In this paper, we investigate these attacks and propose a novel memory architecture for attack resilient IoT network. The information redundancy present in a homogeneous peer-to-peer connected IoT network is exploited to restore the corrupted memory of any IoT node when under attack. We are able to build a failsafe IoT system with STTRAM based program memory which allows guaranteed execution of all the IoT nodes without complete shutdown of any node under attack. Experimental results using commercial IoT boards demonstrate the latency and energy overhead of the attack recovery process.

## I. INTRODUCTION

Internet of Things (IoT) is one of the fastest growing compute segment. It has been predicted that by the year 2020, there will be around 40 billion smart devices connected via the IoT platform [1]. These smart devices will change the way we interact with the environment, thereby spawning a whole array of new application domains like home automation, industrial devices, wearable technology, healthcare monitoring, logistics, to name a few.

IoTs can be of various types and designed for very specific applications. Major semiconductor companies such as Intel, NXP, Qualcomm, RPi Foundation, etc. have come up with their own IoT solutions for smart devices. Various prototyping IoTs are currently available in the market such as Arduino [2], Qualcomm Dragonboard [3], Raspberry Pi [4], etc. The application of IoTs range from home automation such as smart bulbs and automated temperature controllers, wearable technology such as fitness bands and smart watches, healthcare such as medication dispensing systems, industries such as weather and climate monitoring, and, agriculture, machinery and so on. In order to cater to these vast array of application areas, these IoT devices need to be small, fast and energy efficient. Vast majority of IoTs are energy constrained and there is a growing need to reduce the power consumption of these devices [5]. Despite the constraints on size, memory, cost and power, these IoTs are desirable to have long operational lifetime for unattended continuous execution. The IoTs are often used in a distributed network environment [6] and due to the critical nature of the applications, data security and privacy is a growing concern for such a distributed system [7].

IoTs are primarily microcontroller based embedded devices with limited amount of memory. Typically, the application firmware for the IoT is stored in eFlash based memory. However, eFlash suffers from high latency (~us), low endurance (~100,000 Program/Erase cycles) and high write energy (pJ per bit). STTRAM on the other hand has much better latency (~ns), low power consumption (~fJ per bit) and high endurance (~$10^{16}$ cycles), which makes it a promising viable candidate to replace the eFlash based memory on IoTs. However, it has been shown that STTRAM is susceptible to data security and data privacy attacks [8]-[11]. Data security pertains to data corruption by malicious attack with the intention to launch Denial-of-Service (DoS). These attacks exploit the fact that STTRAM is fundamentally susceptible to ambient parameters such as magnetic field and temperature. For STTRAM LLC, tampering during active mode of operation is critical than tampering in power down mode. This is true since the LLC is always invalidated at power ON. However, when STTRAM is employed to store the program (such as eFlash replacement), attacks during both active and power down mode becomes critical. Furthermore, the preventive solutions to maintain the integrity of STTRAM LLC cannot be extended to STTRAM program memory.

Data privacy pertains to sensitive data such as keys and passwords being compromised. Storage such as Hard Disk Drive (HDD) has been the non-volatile part of memory system traditionally protected by encryption. Although effective, the latency associated with encryption makes it non-trivial for application in higher levels of memory stack especially LLC. For eFlash replacement in SoC environment the data security concerns are more serious than data privacy since the data privacy attack models for LLC does not hold true in the proposed application.

In this paper, we investigate possible attack models on STTRAM program memory. We also propose a robust and secure fault-tolerant IoT network architecture which is capable of tolerating magnetic and thermal attacks on embedded STTRAM based program memory. We assume attack sensors such as STTRAM with low free layer volume and weak write [8] to detect the attack. The activities are disabled during the attack and the STTRAM contents are discarded after attack. Finally, the program memory is recovered through peer-to-peer connection from neighboring healthy units. A small portion of the Electrically Programmable Read Only Memory (EPROM) on the IoT is dedicated to store the recovery routine that is executed to restore the programs and applications. The proposed ideas are validated using commercial IoT boards interconnected with network interface card. *To the best of our knowledge this the first effort towards replacing eFlash with STTRAM.*

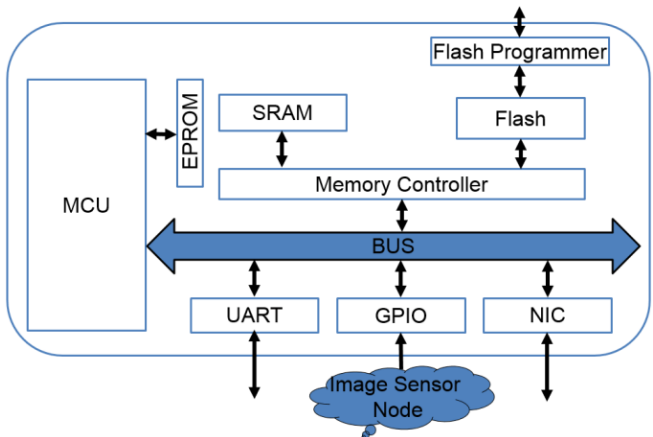

Fig. 1. Simplified structure of a generic IoT with MCU, memory and peripherals.

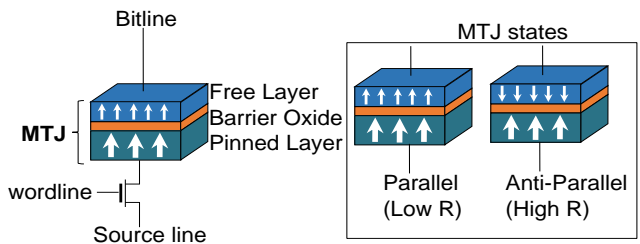

Fig. 2. Schematic of STTRAM bitcell showing MTJ.

TABLE I. Comparison between NAND Flash and STTRAM

| Device Type | NAND Flash | STTRAM |
|---|---|---|
| Present Density | 64 Gb/chip | 2 Mb/chip |
| Cell size (SLC) | $4F^2$ | $4F^2$ |
| MLC Capability | 4 bits/Cell | 4 bits/Cell |
| Program Energy/bit | 10 nJ | 0.02 pJ |
| Access Time (W/R) | 200/20 us | 10/10ns |
| Endurance/Retention | $10^5$/10 yr | $10^{16}$/10 yr |
| Cost/GB | 10x | 1x |

The rest of the paper is organized as follows: Section II provides a background of the existing IoT device architecture, STTRAM basics and the attack model. Section III describes the proposed architecture. Section IV shows experimental results of the proposed architecture. Section V contains some relevant discussions on the proposed work, and conclusions are drawn in Section VI.

## II. BACKGROUND

### A. Overview of a generic IoT device

A generic IoT device has the following basic components as shown in Fig. 1. The Micro-Controller Unit (MCU) is the main logic and controlling system of the IoT. The input/output connections available to the IoT are serial UART (Universal Asynchronous Receiver/Transmitter), GPIO (General Purpose Input and Output), and a NIC (Network Interface Card). The UART is used for serial asynchronous data transfer between devices. The GPIO pins are used to connect sensors, actuators, and other additional components to the IoT. The NIC, either Ethernet or WiFi is used to provide a network identity to the IoT and connect it to a central server or other IoTs in a network.

There are two types of memory devices in the IoT, eFlash and SRAM. The eFlash memory is used to store the application firmware for the IoT. SRAM is used to store dynamic runtime data. The eFlash memory is split into two partitions, a small Bootloader section with lock bits and a larger section for the application program. The eFlash is programmed before the first run of the device using a dedicated flash memory programmer. The available memories on the device are interfaced to the MCU via a memory controller which is responsible for selective memory access from the available range of memory. There is another small low speed memory on the IoT namely the EPROM. It is a low speed electrically programmable hardware, which holds the bootrom that triggers the bootloader code present in the eFlash. Some portion of the eFlash is also made available to store persistent data across reboots.

### B. Overview of STTRAM

Fig. 2 shows the STTRAM cell schematic with Magnetic Tunnel Junction (MTJ) as the storage element. The MTJ contains a free and a pinned magnetic layer. The resistance of the MTJ stack is high (low) if free layer magnetic orientation is anti-parallel (parallel) compared to the fixed layer. The MTJ can be toggled from parallel to anti-parallel (or vice versa) by injecting current from source-line to bitline (or vice versa). The data in MTJ is stored in the form of magnetization. The data stored is '1' if the free layer magnetization is anti-parallel to fixed layer magnetization and '0' if they are parallel.

### C. Benefits of replacing Flash with STTRAM

Table I summarizes the comparison between NAND-flash and STTRAM [12]. The eFlash memory suffers in performance due to its high latency, ~20-200us based on read or write. STTRAM, on the other hand, has a much better read and write performance, close to few nanoseconds. Flash memory also has a lower lifetime as it tends to wear away faster. Furthermore, eFlash requires large write current and is costly. STTRAM is written with a small current and read by evaluating the sense margin and does not wear away like eFlash. STTRAM also has a very low footprint, and can achieve densities as high as DRAM. This makes STTRAM a viable replacement for the eFlash on the resource constrained embedded devices like IoTs.

## D. Attack model: magnetic attack on STTRAM

The magnetization orientation of the pinned layer is fixed using an anti-ferromagnetic coupling and it cannot be changed using nominal current or external magnetic field. Contrary to this, the free layer could be toggled by passing current or by applying magnetic field. The magnetization dynamics of the MTJ free layer is governed by LLG equation [13].

$$\frac{\partial \vec{m}}{\partial t} = \underbrace{-\gamma \vec{m} \times \overrightarrow{H_{eff}} - \alpha\gamma\vec{m} \times \vec{m} \times \overrightarrow{H_{eff}}}_{\text{Field term}} + \underbrace{\frac{I_s \hbar G(\psi)}{2e} \vec{m} \times (\vec{m} \times \overrightarrow{e_p})}_{\text{STT term}} \quad (1)$$

Where $\vec{m}$ is unit vectors representing local magnetic moment, $I_s$ is spin current, $G(\psi)$ is the transmission co-efficient, $\hbar$ is reduced Planck's constant, $\alpha$ is Gilbert damping parameter and $\overrightarrow{e_p}$ is the unit vector along fixed layer magnetization. The effective field is represented by $\overrightarrow{H_{eff}} = \overrightarrow{H_a} + \overrightarrow{H_k} + \overrightarrow{H_d} + \overrightarrow{H_{ex}}$, where $H_a$ is applied field, $H_k$ is anisotropy field, $H_d$ is demagnetization field and $H_{ex}$ is exchange field.

In STTRAM the writing of MTJ is done using STT term (for low power consumption) and external field $H_a$ is kept 0. However, $H_a$ can also be used to toggle the magnetization in absence of charge current (field term, eq. (1)). Note that magnetic field-based toggling is the foundation of Magnetic RAM. The attacker can exploit this extra knob to corrupt the free layer data [8] [9]. Both permanent magnet as well as electromagnet could be used for tampering by the adversary.

It has been noted in [8] [9] that the attacks on STTRAM could be launched either through static (DC) magnetic field or alternating (AC) magnetic field. The DC attack is less detrimental as it can only create unipolar failures. For example, a magnetic field will cause failures only for the bits whose free layer orientation is opposite to the applied field. However, the AC field could cause more damage as it will affect both storage polarities. Due to ease of AC field generation using a low-cost electromagnet this type of attack is highly likely. There are two attack scenarios:

(a) Attack during active mode: The objective of the attack is to launch Denial-of-Service (DoS). A carefully orchestrated DoS attack can result in severe consequences during secure data processing and financial transactions to name a few.

(b) Attack during passive mode: The magnetic attack can also be carried out when the system is OFF. In the proposed application in STTRAM program memory such attacks could boot the system in unwanted state.

## E. Sensor design

The attack can occur while the IoT is powered ON and powered OFF. Therefore, both active and passive sensors should be deployed to sense the attack.

(a) Active sensor: The purpose of the active sensor is to sense the attack few microseconds before the failure of functional bits so as to allow saving of current execution states to EPROM. The sensor sends a programmable interrupt to the MCU when it detects a magnetic attack. The active sensors can be designed

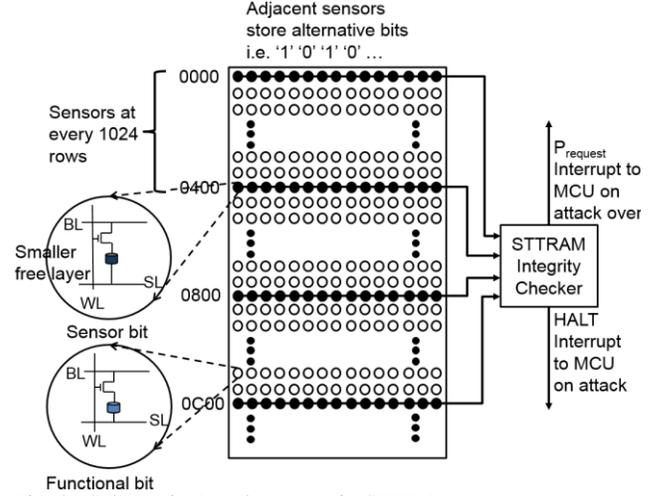

Fig. 3. Schematic Attack sensors in STTRAM array.

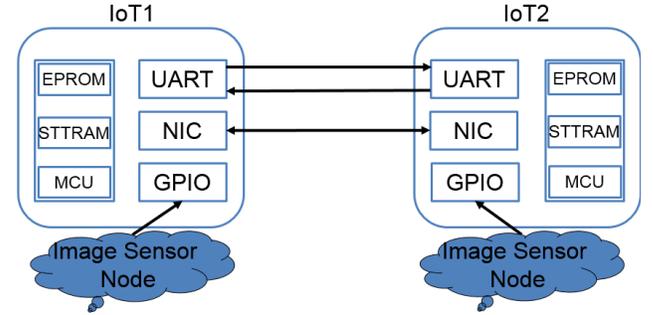

Fig. 4. Dual homogenous IOT network.

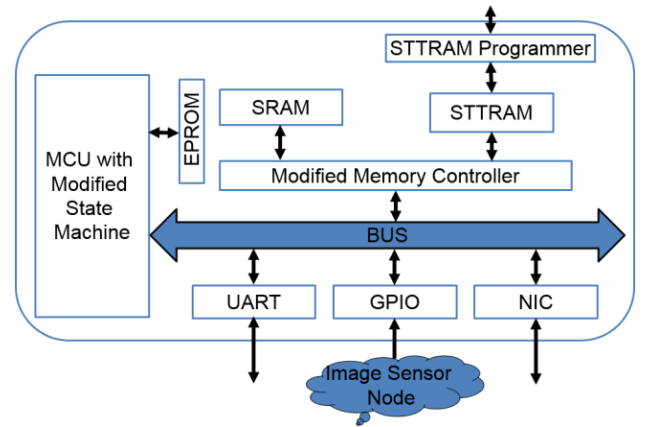

Fig. 5. Modified structure of IoT with STTRAM program memory and modified MCU state machine.

using lower free layer volume and injection of disturb current [8] [9].

(b) Passive sensor: The purpose of the passive sensor is to detect the attack when IoT is powered OFF to prevent the system from consuming tampered program/data when the IoT is powered ON. The passive sensor is essentially same as active sensor without the disturb current.

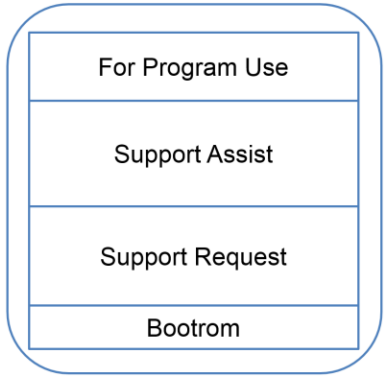

Fig. 6. Modified EPROM layout of IoT.

Fig. 3 shows the deployment of active/passive sensors every 1024 row of the memory array. Adjacent sensors in a row stores '1' and '0' alternatively. This ensures detecting both polarities of magnetic attack. However, the retention time of the sensor should be tailored to be more than the estimated OFF time of the IoTs so that the sensor error in passive mode should be correlated to the attack and not to the retention failures. When the IoT is powered ON, before initializing the application firmware, the sensor addresses are checked by the STTRAM Integrity Checker on the STTRAM Memory Module for signs of any previous attack. If any attack is detected, the IoT goes into the recovery mode to request other IoT on the network to provide the backup firmware.

### III. PROPOSED FRAMEWORK

In this section, we describe the proposed framework. The proposed approach followed by different stages of the IoTs during normal and attack scenarios are presented.

#### A. Proposed approach

For simplicity we consider a dual homogeneous IoT network as shown in Fig. 4. The network consists of two identical IoTs. The eFlash of both the IoTs have identical application programs to perform the same function. The two IoTs are connected to the same network over WiFi or Ethernet interfaced through their respective NIC. The two devices are also connected over a serial communication channel interfaced through their UART ports for special purposes. The required sensors for the IoTs are connected to their respective GPIO ports.

In the proposed design (Fig. 5), the eFlash is completely replaced with STTRAM. However, this poses the possibility of magnetic or thermal attack as discussed above. Since the STTRAM now contains the application program, any attack on the STTRAM scrambles the program binary and results in a complete disruption and failure of the normal sequence of operations in the IoT. There is also no possibility to recover and resume normal operation. The solution is to include some failsafe routine to provide a backup and restore functionality of the program memory. The backup and restore functionality is introduced by adding two special program routines: Support Request and Support Assist in the EPROM. To include this, the EPROM is segmented into four parts, as shown in Fig. 6. The first segment is reserved for the bootrom which is a small write protected segment that is first run when the IoT is powered up. It is responsible for loading the bootloader from the STTRAM. The second and the third segments are reserved for the Support Assist and Support Request programs for the backup and restore routine. The overhead of Support Assist and Support Request code is very low (few bytes). The Support Request code has instructions to send a request message, receive backup firmware bytes, write the bytes to STTRAM and reboot. The Support Assist code has instructions to read firmware bytes from STTRAM, send the bytes to the requesting IoT and reboot. The total space overhead for the two routines is less than 10 bytes (assuming one byte for each instruction). The rest of the EPROM is available to the application firmware to store any persistent data.

The STTRAM memory modules of the IoT devices in the network are equipped with active and passive attack sensors that are capable of sensing any magnetic or thermal attack when it is ramping up (Fig. 3). These sensor arrays are placed at regular intervals of 1024 rows so that they are evenly distributed over the STTRAM. Each sensor array is connected to a STTRAM Integrity Checker on the memory module. The STTRAM Integrity Checker checks the sensor arrays at a periodic interval of a few microseconds for any ramping attack. When any IoT senses an attack it stalls the system. After the attack subsides it executes the Support Request routine to request for memory recovery from the other IoT device on the network. When an IoT receives a memory recovery request from the other IoT on the network, it executes the Support Assist routine to send the

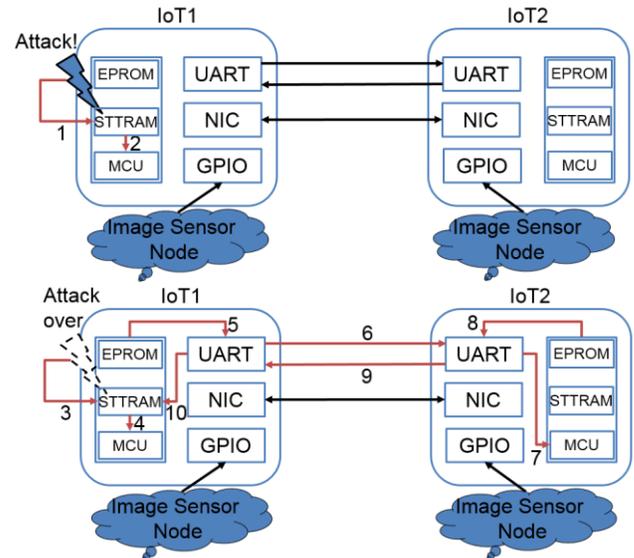

1. Attack sensed by attack sensor · 2. HALT Interrupt · 3. Attack over sensed by attack sensor · 4. $P_{request}$ interrupt · 5. Execute Support Request in EPROM · 6. Recovery Request message · 7. $P_{assist}$ interrupt · 8. Execute Support Assist in EPROM · 9. Firmware recovery data · 10. Write firmware to STTRAM recovery data from its own program memory back to the IoT under attack. The detailed steps performed by the IoT in the attack scenario (Fig. 7) are described below.

Fig. 7. Attack handling. The sequence of events are numbered and explained.

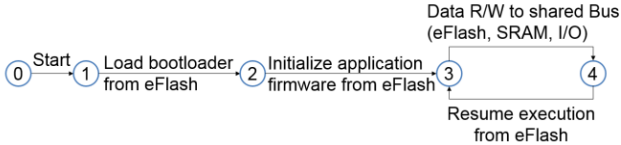

Fig. 8. State machine of the MCU of a generic IoT.

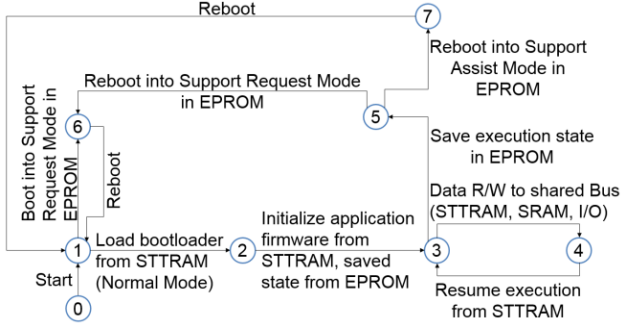

Fig. 9. Modified state machine of the proposed architecture.

### B. Boot up and normal operation

When the IoT is powered up, the bootrom in the EPROM starts the hardware tests which runs the STTRAM Integrity Checker on the memory module to check for the authenticity of the STTRAM program memory. After passing the integrity test, the bootrom triggers the bootloader code in the STTRAM. The bootloader in the STTRAM initializes the application firmware from the Program Memory. It looks for any saved execution state in the EPROM and if present, loads the MCU with the saved state to resume execution form that state. The saved state in the EPROM is erased to prevent any inconsistent state configuration of the MCU for future reboots. If no saved state is present, the MCU starts with a clean state. During normal operation the IoT sensors (e.g, image sensor nodes) capture data and send the data to be processed to the IoT. The data is transferred to the IoT through the GPIO ports and the memory controller transfers it to the SRAM to be processed by the MCU. The IoT can process the data or store it to the STTRAM. It can also share the data to the other IoT in the network by transmitting through the network interface.

### C. Attack sensing

During the normal operation of the IoT, if an adversary tries to attack the STTRAM with the intention to scramble the stored firmware, the attack sensors and the Integrity Checker is able to sense the attack ahead of time. The sensor array is preconfigured to hold an alternate sequence of '0' and '1'. In case of an attack the sensor array on the STTRAM no longer retains their original sequence. The STTRAM Integrity Checker detects the scrambled sensor arrays and sends the HALT interrupt as an attack signal to the microcontroller. When the microcontroller receives the HALT interrupt, it halts the running application and saves the current execution state to the EPROM section available for program use. This ensures that no data is lost. Moreover, this allows the system to resume normal operation from the exact same state when the attack is over. When the attack subsides, the attack sensors senses the magnetic field strength going down, and removes the HALT interrupt from the microcontroller. The Integrity Checker then sends a programmable interrupt, $P_{request}$ to the microcontroller which reboots the IoT in Support Request mode and starts executing the recovery request code from the EPROM.

If the adversary tries to launch the attack when the IoT is powered OFF, the passive sensors are able to detect the attack due to failure of sensor bits. When the IoT is powered up after the attack, the bootrom triggers the STTRAM Integrity Checker and the STTRAM will fail the integrity check due to the modified sensor array from the previous attack. The bootrom then sets the working mode of the IoT to the Support Request mode and starts executing the recovery request code from the EPROM.

### D. Support request

In the Support Request mode, the IoT starts executing the recovery request code from the EPROM. The recovery request routine writes a request code message at the UART Tx port to send to the other IoT on the network. Since this is a critical operation, a wired connection is preferable to a wireless network interface. The request routine after sending the request code message waits to receive data from the UART Rx port. The Support Request code consists of the following instructions:

1. Send recovery request message
2. Receive backup firmware byte
3. Write firmware byte to STTRAM address
4. Increment STTRAM address index
5. Reboot

Assuming 1 byte for each of the instructions, the space overhead of Support Request routine is 5 bytes.

### E. Support assist

The IoT microcontroller is programmed to listen to a programmable interrupt, $P_{assist}$. When an IoT receives the request code message on its UART Rx port, the UART controller sends a programmable interrupt $P_{assist}$ to the microcontroller. On receiving the interrupt, the microcontroller saves the current execution state to the EPROM, and reboots the IoT in Support Assist mode and starts executing the recovery assist code in the EPROM. The recovery assist routine starts reading the application firmware in the program memory and bootloader code from the STTRAM and writes serially to the UART Tx port to send to the requesting IoT on the network. The Support Assist code consists of the following instructions:

1. Read firmware byte from STTRAM address
2. Send firmware byte to UART Tx
3. Increment STTRAM address index
4. Reboot

Assuming 1 byte for each of the instructions, the space overhead of Support Assist routine is 4 bytes.

*F. Recovery*

When the assisting IoT starts sending the firmware over the UART connection, the recovery request routine writes the Program Memory and Bootloader partitions with the new firmware, overwriting any potential scrambled code from the attack. When the firmware transmission is complete, the sensor arrays are reset to its original configuration of alternating '0' and '1'. The recovery assist routine on the assisting IoT after transmitting the entire firmware from the STTRAM, reboots the IoT in normal operation mode.

On a normal IoT device, a hypothetical state machine of the MCU can have four states (Fig. 8). When the device is powered off, it is in State-0. When the device powers up the MCU is in State-1, where it bootloader from EPROM loads the bootloader from eFlash (State-2). When the bootloader initializes the application firmware from the eFlash, it goes to State-3, where it starts executing the program. For any data read or write operation on the shared bus (connected to the eFlash, SRAM, I/O etc.) it goes to State-4. When it resumes execution from eFlash, it goes back to State-3.

In the proposed framework, the state machine is modified to include the other backup and restore functionality (Fig. 9). The States from 1 to 4 remain same as before. After power-on, the STTRAM Integrity Checker runs the test. If the test fails, it sets the operation mode of the device to Support Request mode (State-6) and starts executing the recovery request routine. After recovery completion, it reboots to State-1. When running in normal operation mode, if the execution state needs to be saved to EPROM to start the Support Request/Assist routine, it goes to State-5. It then reboots the device in Support Request Mode (State-6) or Support Assist mode (State-7). When executing the Support Request routine from EPROM, it receives the backup firmware and writes to STTRAM. After the recovery is complete, it reboots the device in normal operation mode (State-1). When executing the Support Assist routine from EPROM, it reads the firmware from STTRAM and transfers to the requesting IoT. After the transfer is over, it reboots the device in normal operation mode (State-1).

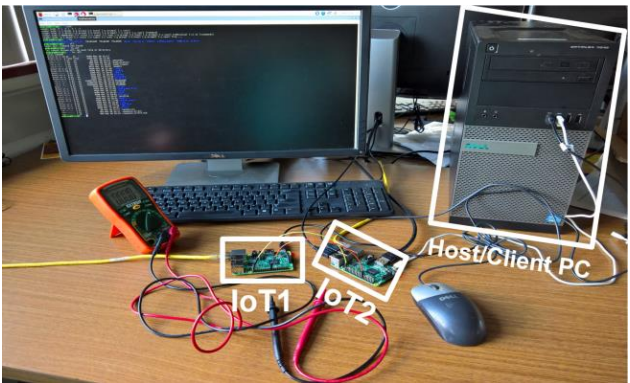

Fig. 10. Experimental setup. Two IoTs are connected using Ethernet/WiFi. IoT1 is under attack whereas IoT2 provides the support.

TABLE II. Specifications of DragonBoard and Raspberry Pi

| IoT | DragonBoard 410c | Raspberry Pi 3 |
|---|---|---|
| SoC | Qualcomm Snapdragon 410 | Broadcom BCM2837 |
| CPU | Quad-core ARM® Cortex® A53 at up to 1.2 GHz per core | 4× ARM Cortex-A53, 1.2GHz |
| GPU | Qualcomm Adreno 306 | Broadcom VideoCore IV |
| RAM | 1GB LPDDR3 (533MHz) | 1GB LPDDR2 (900 MHz) |
| Networking | Wi-Fi 802.11 b/g/n 2.4GHz | 10/100 Ethernet, 2.4GHz 802.11n wireless |
| Bluetooth | Bluetooth 4.1 | Bluetooth 4.1, BLE |
| Storage | 8GB eMMC 4.5 / microSD 3.0 | microSD |
| GPIO | 40-pin header | 40-pin header |

TABLE III. Experimental results for 100MB data transfer

| Network | Ethernet | WiFi |
|---|---|---|
| Latency | 9 sec | 1023 sec |
| Average data rate | 89 Mbps | 782 Kbps |
| Extra current drawn | 0.37A | 0.35A |
| Energy/bit | 8.44 nJ | 190 nJ |

IV. EXPERIMENTAL RESULTS

The energy and latency overhead of the proposed recovery mechanism is estimated using a network of Qualcomm DragonBoard 410c and Raspberry Pi3 IoT boards (Fig. 10). The detailed specifications of the IoTs used are given in Table II [3] [4]. The IoT devices are tested with both wired and wireless network connections. The devices are powered with a regulated DC voltage source of 5V and the power drawn on each board in idle situation is observed to be 0.32A. The attack scenario is triggered with a user input to the software running on the IoT1, which triggers the transfer of 100 MB data from the supporting IoT (IoT2). The power consumption statistics is shown in Table 3. From Table III we can conclude that energy consumed to restore the program memory of size 100MB using Ethernet (WiFi) connection is 8.44J (190J). The corresponding latency overhead is 9sec (1023sec). The energy and latency overhead could be minimized by reducing the amount of recovery data. This can be achieved by using STTRAM with higher energy barrier to make them robust against magnetic field, implementing stronger ECC to protect against low error rates, identifying the corrupted segments of the program memory using ECC and fetching only those blocks from healthy units. Future research will focus on developing a simulation framework to quantify the tradeoff of replacing eFlash with STTRAM with respect to attack resilience.

## V. DISCUSSIONS

### A. Authenticity and privacy of received data

An adversary can cause authenticity and privacy issues during the recovery process by tampering the communication between two IoTs. The issues are as follows: (a) The adversary can snoop the data during transfer thus getting access to sensitive information; (b) The adversary can also inject tampered data during the transfer; and, (c) The adversary can mimic the support request. The IoT that receives the request may not recognize that the request is from an adversary and not an IoT; thereby sending the sensitive program data to adversary. To avoid these scenarios, data can be encrypted using public-key encryption. The corresponding public-private keys should be stored in the EPROM of the IoTs which adds some storage overhead. Physically Unclonable Functions (PUFs) [14] [15] can also be employed to generate keys.

### B. Issues related to mode of communication

The communication between two IoTs can be either wired or wireless. However, these two modes have some pros and cons. The wired connection is faster, more reliable and less prone to interference compared to wireless communication. However, over the long distance the signal strength drops and the wired connection can also be physically tampered by the adversary thus data security is breached. On the other hand, Bluetooth has its own embedded data protection. However, Bluetooth has very short communication distance and again, adversary can hamper any wireless communication using a physical barrier. Therefore, depending upon the requirement mode of operation can be selected.

### C. Other modes of attack

Although this paper focused on magnetic attack, STTRAM is also susceptible to thermal attack. The proposed approach is equally applicable to thermal attack scenarios. A temperature sensor based on STTRAM bitcell can be employed to detect the attack and trigger the recovery procedure proposed in this paper.

### D. Applicability to other non-volatile memory technologies

The proposed methodology could also be extended to evaluate the feasibility of replacing eFlash with other emerging non-volatile memory technologies such as Resistive RAM (ReRAM) and Phase Change Memory (PCM) [12].

## VI. CONCLUSION

In this paper the implications and challenges of replacing eFlash with STTRAM in embedded devices like IoTs are discussed. A novel attack resilient architecture is proposed which allows the devices on the network to recover from attacks and continue execution without shutting down completely. The energy and latency overheads of the proposed architecture is presented through the experimental results.